*Research Article*

# Multitier Service Migration Framework Based on Mobility Prediction in Mobile Edge Computing

**Run Yang, Hui He, and Weizhe Zhang**

*School of Computer Science and Technology, Harbin Institute of Technology, Harbin, China*

Correspondence should be addressed to Hui He; hehui@hit.edu.cn

Mobile edge computing (MEC) pushes computing resources to the edge of the network and distributes them at the edge of the mobile network. Offloading computing tasks to the edge instead of the cloud can reduce computing latency and backhaul load simultaneously. However, new challenges incurred by user mobility and limited coverage of MEC server service arise. Services should be dynamically migrated between multiple MEC servers to maintain service performance due to user movement. Tackling this problem is nontrivial because it is arduous to predict user movement, and service migration will generate service interruptions and redundant network traffic. Service interruption time must be minimized, and redundant network traffic should be reduced to ensure service quality. In this paper, the container live migration technology based on prediction is studied, and an online prediction method based on map data that does not rely on prior knowledge such as user trajectories is proposed to address this challenge in terms of mobility prediction accuracy. A multitier framework and scheduling algorithm are designed to select MEC servers according to moving speeds of users and latency requirements of offloading tasks to reduce redundant network traffic. Based on the map of Beijing, extensive experiments are conducted using simulation platforms and real-world data trace. Experimental results show that our online prediction methods perform better than the common strategy. Our system reduces network traffic by 65% while meeting task delay requirements. Moreover, it can flexibly respond to changes in the user's moving speed and environment to ensure the stability of offload service.

## 1. Introduction

With the development of smart devices, more applications such as artificial intelligence, autonomous vehicles, interactive gaming, virtual reality, augmented reality, and smart surveillance systems demand intensive computation resources and high energy consumption for real-time processing [1]. However, smart devices with limited computation and energy capacity cannot effectively support the applications due to the constraint of physical size. The traditional approach is to cooperate with remote cloud centers. This approach has a relatively large transmission delay, which brings a large burden to the core network, and it is not suitable for low-latency applications. Mobile edge computing (MEC), proposed by the European Telecommunications Standards Institute in 2014, provides computing resources for the user equipment (UE) at the edge of the network to reduce the computational delay and energy consumption of the UEs and address this challenge [2, 3]. MEC can improve quality of service (QoS) and quality of experience (QoE) [4–10]. Concepts similar to MEC are cloudlet [11], edge cloud [12], Fog Computing [13], and Follow me Cloud [14].

Users should offload tasks to a nearby MEC server to ensure low latency. However, tasks will be offloaded to different MEC servers due to user mobility and limited coverage of base stations or Wi-Fi hotspots [15]. When users leave their original location, their services must be migrated to a new MEC server close to the current users' location. An efficient service migration scheme should be employed in the network edge to realize seamless service migration (i.e., without disruption of ongoing edge services) and ensure service continuity when users move between the service ranges of different MEC servers. When an arbitrary user moves out of the service region of the relevant MEC server, whether and where to migrate the edge services in progress should be decided [16].

Furthermore, service migration includes stateful migration and stateless migration. Most popular interactive applications need to consider the context of the application. Thus, the internal state of the application needs to be kept for the user. Virtualization techniques are used for service migration, such as virtual machine (VM) and containers to address this challenge. However, only the runtime environment of running applications is stored due to the storage resource constraints of the MEC server. When service migration occurs, the image of the runtime environment needs to be copied to the target MEC server and destroyed after use. Image sizes range from a few hundred megabytes to a few gigabytes [17]. These files can be transferred in batches, and most of the data can be transferred before the service handoff, reducing service downtime [12, 18].

Therefore, predictive migration [19] should be considered to ensure the seamless migration of services. However, existing research works are based on prior knowledge, such as historical data of users [20–25]. A large amount of user data needs to be collected for analysis, and a long adaptation period is required to achieve satisfactory prediction accuracy. These methods have the following shortcomings: (1) scalability is poor, cannot adapt to the new environment and new users; (2) universality is poor, need to model for each user; (3) data management is complex, the need to collect and store a large number of user privacy data, but also to ensure the security of data; (4) these methods are complex, require a large amount of computing and storage resources.

Moreover, the latency requirements of offloading tasks and the speed of the user's movement during service migration are not taken into account by the existing research works. Low-latency tasks should be offloaded to MEC servers as close to the data source as possible to ensure a low transmission delay. In MEC environment, deploying MEC servers and base stations together can address this challenge, but base stations have limited service coverage, and because of user mobility, users will connect to different base stations and generate service migrations. The high-speed movement will produce frequent service migration and service interruption that will generate a large amount of network traffic, which leads to edge network overload and service instability.

To overcome the dependence on prior knowledge, an online real-time movement prediction method is proposed. We investigate service migration with digital maps. Instead of relying on user history information for prediction, we explore road maps and the user's position to predict the user's movement. When users travel in a city, their movements are not random, and they have to walk along roads. Figure 1 shows a practical scenario when a user is moving on the road. Only two directions are available: forward and backward. Thus, which MEC server coverage they will enter in the future can be predicted. However, with the advancement of science and technology, digital maps and positioning technology [26] have been well developed with high precision and have played an important role in our daily life (i.e., Baidu map and Amap). Combined with the digital map and MEC server distribution map as well as the real-time location information and movement direction of users, the next MEC server can be accurately predicted.

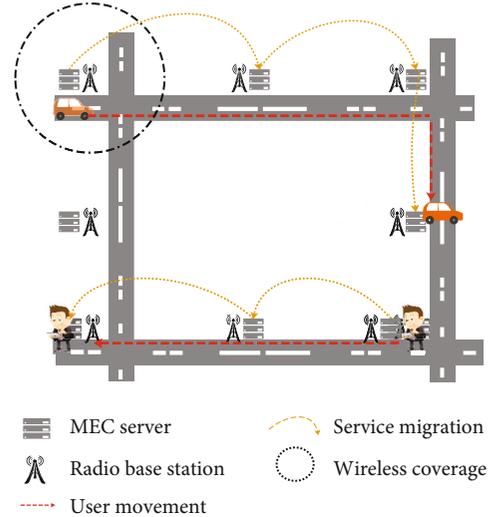

FIGURE 1: Migration scenarios in MEC.

To address the challenge of frequent migration caused by high-speed movement, we take into account the latency requirements of offloading tasks and the speed of the user's movement. A multitier service migration framework (MSMF) is proposed, divided according to the service coverage of the MEC server. The underlying MEC servers are deployed with base stations and are used for low-latency tasks. The upper MEC servers are connected to multiple base stations, which increase the coverage several times and also increase the transmission delay, and will be used by tasks that can tolerate high delay. The tradeoff between task delay requirements and the consumption generated by service migration should be made. Under the premise of meeting the delay requirements, the upper level edge servers should be selected as far as possible, and the lower level edge servers with limited resources should be reserved for tasks requiring extremely low delay. Reducing the number of service migration can guarantee the stability of service and improve the quality of user experience. Service migration takes into account the types of offloading tasks and the service deployment framework, which will achieve better performance [27].

MSMF is a highly extensible framework based on online prediction, does not need to pay attention to the prior knowledge of each user, and is not affected by environmental changes. This work is the first where service migration is based on user movement speed, task latency attributes, and digital maps to the best of our knowledge.

Our contributions to this paper are summarized as follows:

(i) We propose an online prediction method based on digital maps. A mapping relationship between the deployment location of MEC servers and road data is established, and the user's movement status (position and direction) is obtained in real time to predict the next access to the MEC server

(ii) We propose a multitier framework based on MEC service coverage. The appropriate MEC server is

selected according to the movement speed and delay requirement of the offloading task. While meeting the delay requirements of offloading tasks, network load is minimized, and QoS and QoE are improved

(iii) Extensive experiments are conducted to validate the proposed method. The simulation platform is used for experiments, and a physical environment is built to conduct experiments based on real user movement trajectory data

The rest of this paper is organized as follows. Section 2 reviews the related work. Section 3 presents the system overview. Section 4 proposes the prediction method of service migration and the migration algorithm of MEC server deployment based on a multilayer architecture. Section 5 presents the experimental evaluation. Section 6 concludes this work.

## 2. Related Work

In this section, the live migration technology used in the service migration, the research work related to service migration in MEC, and the prediction-based service migration research work in MEC are introduced.

*2.1. Live Migration.* Migration techniques used for service migration include cold migration and hot migration [28]. Cold migration freezes/stops the service before starting to copy its state to the target server, copying all data to the target server at once. This approach can result in substantial service downtime. The live migration approach keeps the service running, while most of its data is being transferred to the target. The service only pauses when the minimum amount of overall status is transmitted, after which the service runs on the new server.

Precopy migration copies most of the data before freezing the service for a final minimal data transfer (i.e., memory data). It performs service migration based on iterative transmission, and each iteration updates the target server with the latest data. The first iteration copies all the data (as cold migration does). During the data transfer in the first iteration, the executing service may modify the memory data. The modified memory pages are called "dirty pages," and the next iteration only transmits them to update the target server. Typically, the iterations of the precopy phase are convergent, that is, the duration becomes increasingly shorter. The downtime for precopy migration is usually short. This technology is excellent for service migration in MEC. Regarding containers, Checkpoint/Restore In Userspace (CRIU) [29] provides all basic mechanisms (i.e., CRIU predump function) necessary for precopying the runtime state of the service. However, it does not provide any way to run them as a whole easily. Haul [30] has been solving this issue. It is a software built on CRIU (first implemented as a Python script and then converted to Go) to provide a simple, transparent tool for precopy migration.

*2.2. Migration in MEC.* Several methods for live migration are proposed. These methods are classified according to their basic virtualization technologies, such as VM-based migration, container-based migration, and process-based migration. In [17], Ha et al. proposed a VM handoff technology that can seamlessly transfer the calculation of VM packaging to the optimal unloading site when the user moves. In [31], Hu et al. proposed a dynamic service migration strategy with energy optimization that transforms the migration energy optimization of dynamic services into a migration path selection problem. In [28], Zhang et al. analyzed in detail the VM migration work in the cloud scenario. VM migration includes memory data migration and storage data migration and can be performed using precopy, postcopy, and hybrid copy. In [12], Machen et al. proposed a layered framework for service migration that supports container and VM technologies. The framework decomposes the application into multiple layers and only transfers the missing layers to the destination. This layered approach can substantially reduce service downtime. Ma et al. [18] proposed an efficient migration method for sharing storage files based on the hierarchical nature of the container file system. The transfer size during the migration can be further reduced by transferring the basic image before switching and transferring the incremental memory difference only at the beginning of the migration. According to the above research, service migration can be divided into multiple steps. Data in the service migration can be classified into static data (i.e., image) and dynamic data (i.e., in-memory data). Static data can be transferred ahead of the migration, and dynamic data are only transferred during the handoff, considerably reducing the migration downtime compared with transferring all data during the handoff.

*2.3. Prediction-Based Migration.* In the research work of service migration based on prediction, Wang et al. [23] showed that service migration can rely on a separate prediction module to predict the future costs of each user running its service in each possible MEC server. In [19], a migration mechanism based on user mobility prediction using low-level channel information was proposed. In [32], the Markov mobility model was first used to study the performance of MEC in the presence of user mobility, but the paper did not consider whether and where to migrate the service. In [22, 24], MDP models for state transitions during service migration were defined for 1D and 2D scenarios. The service migration models use value or strategic iterative solutions to design appropriate migration decisions and balance migration costs and user experience. The migration policy includes many factors. In [22], Wang et al. used a 1D asymmetric random migration model and an improved iterative algorithm to find the optimal migration threshold, which can reduce the migration time. However, user mobility is not limited to one dimension in the actual scenario. Thus, in [24], Wang et al. considered a general cost model, 2D user mobility, and only the distance between the user and MEC server locations. Distance-based MDP helps develop effective algorithms and find an optimal policy. These studies analyzed service migration in theory, but several limitations remain in their application to practical scenarios. Wu et al. [25] proposed a factor graph learning model based on some prior knowledge, such as users' social network information, and the correlations between users' locations.

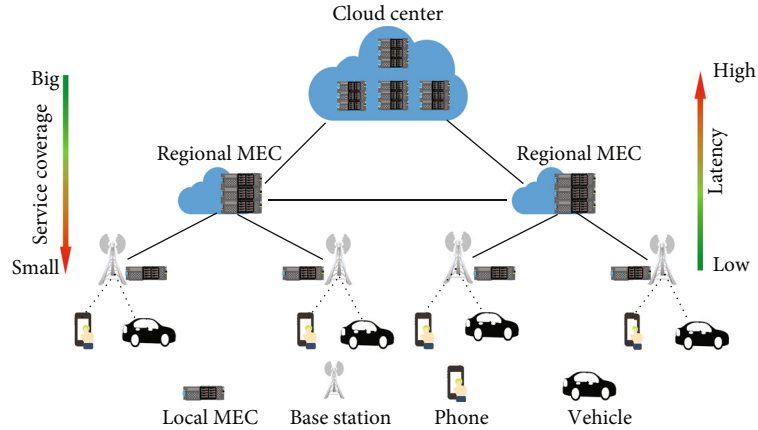

Figure 2: Architecture of MSMF.

## 3. System Overview

In this section, the framework MSMF and the detailed workflow of the framework are introduced. Figure 2 provides an overview of our framework MSMF. The framework can be divided into a control plane and a data plane. Most of the data generated by the service migration is concentrated in the data panel, and only a small part of the data is transmitted to the cloud for analysis and scheduling, which can reduce the load on the core network. MSMF consists of three main components. (1) The cloud component is used to analyze data and schedule migration. (2) The MEC platform is used to run the services. (3) The UE is used to run the application and monitor the data (such as position and network speed) sent to the cloud center

*3.1. Cloud Center.* The cloud center has sufficient computing and storage resources and can be used to store map data and image data, data analysis, and scheduling. Moreover, it is connected to all MEC servers and can observe the status of all MEC servers from a global perspective.

*3.1.1. Map Data.* Map data include a digital map and a MEC server distribution map. The digital map is used to predict the users' trajectories. The MEC server distribution map is used to combine the digital map data to establish the mapping relationship (RMR) between the MEC server and the roads and then use it to predict which MEC server the user will access next based on the status information collected from the UEs.

*3.1.2. Prediction Module.* First, status information such as the user's location, moving direction, movement speed, network delay, and delay requirements for offloading tasks are obtained from the UE. Then, the location data are matched with digital map data to determine the road the user is currently on. Lastly, RMR and moving direction can predict the MEC server to be accessed in the future.

*3.1.3. Scheduling Module.* The cloud center monitors the computing resources and storage resources of each MEC server in real time. When service migration is initiated, the cloud center selects the appropriate MEC server for migration based on several important parameters, such as delay requirement of the offloading task, user movement speed, network delay, and the remaining amount of resources of MEC servers. After the MEC server is selected, a new IP address will be obtained after the running environment is deployed on the new MEC server. The cloud center will notify the UE of the new IP address. After the migration is completed, the offloading task will be sent to the new IP address.

*3.1.4. Image Set.* The storage resource of MEC servers is limited. The runtime environment used when the user offloads the task is made into an image file and stored in the cloud center. When the user initiates the task offloading, the MEC server directly downloads it from the cloud center.

*3.2. Mobile Edge Computing Platform.* MEC platforms can be divided into two categories according to their coverage, namely, local MEC servers and regional MEC servers, and the architecture is shown in Figure 2. Local MEC servers are deployed with radio base stations, and their range of service is the coverage of the base station. Regional MEC servers are connected to multiple base stations, and their range of service is several times that of local MEC servers, but their transmission delay is larger. This system is designed to address the challenge of frequent migration when users move at a high speed. A MEC platform can deploy any operating system and virtual platform, such as containers, while an automated migration service is deployed.

*3.2.1. Virtual Platform.* Virtual platforms include VMs and containers. The VM includes hardware virtualization, and the container is only virtualized at the operating system level, such that its image files are much larger than the containers. Considering the size of the migration data, containers are more suitable for service migration.

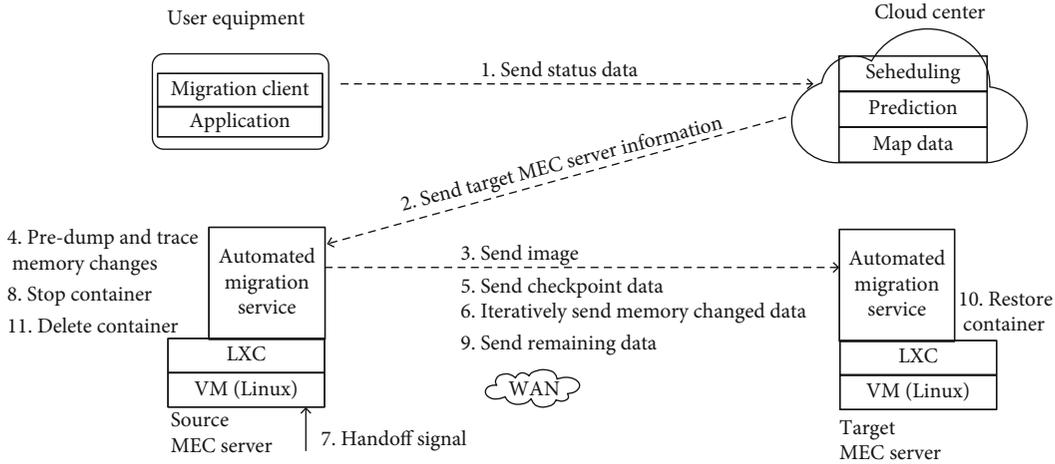

Figure 3: Full workflow of service migration.

*3.2.2. Automated Migration Service.* Automated migration service includes a container migration controller and file synchronization management.

The container migration controller is the core component of the MEC platform, which is used to manage the life cycle of containers, checkpoint the container state, restore the container with checkpoint files, and track the container memory changes.

File synchronization management is used to transfer files and ensure that the files on the source MEC server and the target MEC server are completely consistent. The container data includes container checkpoint files and container image files. These files are static and can be synchronized once and then iterated with CRIU to synchronize the memory data, synchronizing the dirty memory data each time, thus reducing the size of the transferred data.

*3.3. User Equipment.* Two applications are running on the UE. The first is the migration client, and the second is the application that offloads the service to the MEC server.

*3.3.1. Migration Client.* The client consists of two modules that are responsible for monitoring location, movement status, network configuration, network communication quality monitoring, and offloading task delay attribute acquisition.

The status monitoring module tracks the changes in the user's position, moving speed, and moving direction in real time. A policy is set up to send data to the cloud center only when the movement direction and movement distance exceed the set threshold or movement speed changes to avoid the large amount of redundant data generated by frequent interaction between the UE and cloud center. The module also monitors the delay attribute of the task, which determines whether the task is offloaded to the local MEC server or the regional MEC server.

The network module is used to receive the new MEC server IP address from the cloud center and configure and monitor the quality of the network. When the service is migrated from one MEC server to another, it is assigned a new IP address, and the offloaded task needs to be transferred to the new IP address.

*3.3.2. Application.* The application runs on UEs, connects to the MEC server, transmits task data, and receives the results when offloading service occurs.

*3.4. Workflow of Service Migration.* Figure 3 details the migration steps of the designed framework. The user is currently in the service area of the source MEC server and offloads tasks to run on it. The user is about to move to the service area of the target MEC server. Thus, the service needs to be migrated from the source MEC server to the target MEC server. The migration steps are described below.

*s1: Send Status Data.* The client monitor module sends the status data of the user (i.e., position, speed, and direction) and the delay attribute of the task to the cloud center. The cloud center uses this data to predict the next MEC server in real time.

*s2: Send Target MEC Server Information.* The cloud center sends the predicted MEC server information to the current MEC server to prepare for migration. Based on the prediction method proposed, one or more MEC servers may be accurately predicted.

*s3: Send Image Data.* The image from the source MEC server is synchronized to the target MEC server. In addition to the target MEC server, the image can be synchronized to other predicted MEC servers in advance when the network load is low, which can avoid network congestion caused by service migration and affect the quality of network service.

*s4: Predump and Trace Memory Changes.* Memory snapshots are saved while keeping the container running, generating checkpoint files, and enabling memory change tracking.

*s5: Send Checkpoint Data.* The checkpoint files to the target MEC server are synchronized.

*s6: Iteratively Send Memory Changed Data.* The container on the source MEC server keeps running, and the memory data change, which is called dirty memory data. These dirty memory data are periodically synchronized to the target MEC server until the container stops.

*s7: Handoff Signal.* When the user switches from the current base station to another base station, service interruption is triggered, and the automated migration service completes the remaining service migration steps.

*s8: Stop Container.* Once the system receives the switch signal, the container on the source MEC server stops.

*s9: Send Remaining Data.* After the container is stopped, the remaining dirty memory data and the latest files are synchronized to the target MEC server.

*s10: Restore Container.* After the target MEC server receives the latest memory data, the container can be restored with the image and checkpoint data. The service is successfully migrated from the source MEC server to the target MEC server.

*s11: Delete Container.* Containers that do not provide services should be deleted to ensure that other services can be migrated to the current MEC server and provide services due to the limited resources of MEC servers.

## 4. Algorithms

This section introduces map-based prediction methods, multitier framework, and scheduling algorithms. The key notations are described in Table 1.

*4.1. Prediction-Based Service Migration.* Accurately predicting the MEC server that users will visit in the future is essential for service migration based on prediction. For a given container to be migrated, the total migration time includes precopy time and service downtime. Therefore, total migration time is defined as follows:

$$\text{tmt} = \frac{S_{\text{con}} + M + \sum_{i=1}^{N} M_d + M_{\text{handoff}}}{W_{\text{edge}}}, \quad (1)$$

where $N$ is the number of iterations transmitted. The objective is to minimize system migration time (tmt). The prediction-based migration approach completes the data transmission of $S_{\text{con}}$, $M$, and $\sum_{i=1}^{N} M_d$ while the service is still running, and only $M_{\text{handoff}}$ influences service downtime. A threshold $M_d$ is set for iteration transmission of dirty memory to minimize service downtime. Accurate prediction methods can help achieve the goal of completing $S_{\text{con}}$ and $\sum_{i=1}^{N} M_d$ transmission before handoff. Thus, a service migration method is proposed based on map prediction.

The relationship ($MR_{\text{map}}$) between MEC servers (the local MEC servers are colocated with the base station) and roads must first be established to implement map-based service migration. Figure 4 shows that considering a hexagonal cell structure, a road passes through 14: return $S$ and $N$ the coverage area of one or more base stations. The car will pass Cell 1, Cell 2, and Cell 3, and the next MEC server can be determined accurately based on the direction the user moves. When multiple MEC servers to be accessed in the future have been determined, nondata can be synchronized to them in advance when network resources are sufficient, and only a small amount of state data is synchronized when switching between different MEC servers to avoid network congestion and improve QoS. The container can be deployed to Cell 2 and Cell 3 ahead of time, and the dirty memory data are synchronized only during service handoff.

TABLE 1: Key notations.

| Notations | Description |
|---|---|
| $MR_{\text{map}}$ | Map based on road information and MEC server deployment location |
| $R_{\text{map}}$ | Digital map information |
| $V_i$ | Movement speed of user $i$ |
| $L_i$ | Level of latency at which user $i$ offloads task |
| $P_i$ | Current location of user $i$ |
| $D_i$ | Direction of the movement of user $i$ |
| $\text{Dis}_j$ | Coverage distance of MEC server $j$ on the current road |
| $S_{\text{con}}$ | Size of container image |
| $W_{\text{edge}}$ | Bandwidth between two MEC servers |
| $M$ | Size of the first memory data snapshot |
| $M_d$ | Threshold for iterative synchronization of dirty memory data |
| $M_{\text{handoff}}$ | Size of dirty memory during handoff |
| $d_{\text{local}}$ | Network latency at the local MEC server |
| $d_{\text{regional}}$ | Network latency at the regional MEC server |
| $d_p$ | Computing delay at a MEC server |
| $s_{\text{handoff}}$ | Signal for switching wireless network |
| tmt | Total migration time: sum of precopy time and service downtime |

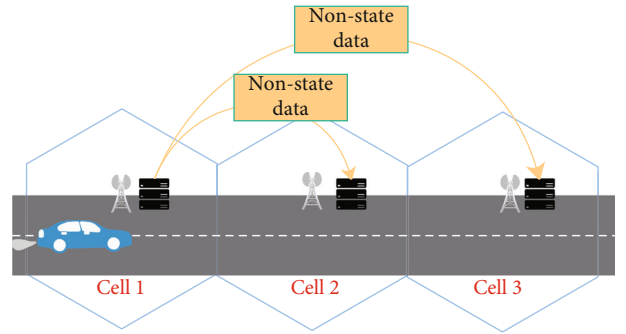

FIGURE 4: Mapping relationship between MEC servers and roads.

However, the coverage of base stations is not evenly distributed on the road, and different base stations cover the same road with different distances ($\text{Dis}_j$). The short-distance coverage will cause frequent service migration and service interruption, resulting in service instability and reducing service quality. For applications in scenarios such as autonomous driving, service interruption should be avoided as much as possible. Otherwise, it will have serious consequences.

Figure 5 shows that the road passes through the coverage area of three base stations, and the distance through the coverage area of Cell 1 is very short. When users move quickly, they will pass through this area very soon. Thus, the tradeoff must consider whether to migrate services to

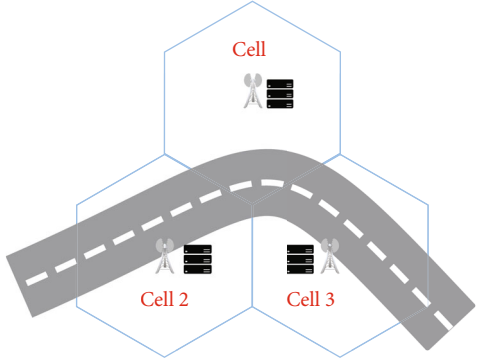

Figure 5: The distance of coverage affects migration decisions.

Cell 1. Considering the application service latency requirements and to meet the service delay requirements, the data in this area are forwarded to Cell 2 or Cell 3 through the base station of Cell 1. Such optimization can reduce the number of service migrations, reduce the redundant traffic generated by service migrations, and improve QoS and QoE.

*4.2. Service Migration Based on Multitier Framework.* Each layer in the multitier architecture proposed in our paper corresponds to the delay and service range of different sizes, as shown in Figure 2, and the size of service range and transmission delay is increased from the lower layer to the upper layer in turn. Low-latency tasks need to be offloaded to the edge servers as close as possible to the data source for processing, to ensure that the task has low transmission latency. In a mobile edge computing environment, deploying edge computing servers and base stations together can address this challenge, but the service coverage of base stations is limited. For users with fast-moving speed, if the underlying edge server is selected, the time that the user resides in the service scope of the edge servers is too short, which will result in frequent migration. Service migration takes some time, and if the service migration is not completed before the user arrives at the new service scope, there will be a long service interruption. Frequent service migration leads to a large number of service interruptions and network traffic, which seriously affects the quality of service and experience. Therefore, we take the delay requirements of offloading tasks and the moving speed of users into account in the service migration and propose a multitier service deployment architecture, which is divided according to the service coverage of the edge servers. The bottom edge servers are deployed together with the base stations and are used by the low-latency tasks. The upper edge servers connect multiple base stations, which increase the coverage several times and also increase the transmission delay, and will be used by tasks that can tolerate high delay. The tradeoff between task delay requirements and the consumption generated by service migration should be made. Under the premise of meeting the delay requirements, the upper level edge servers should be selected as far as possible, and the lower level edge servers with limited resources should be reserved for tasks requiring extremely low delay.

Reducing the number of service migration can guarantee the stability of service and improve the QoS.

The complete procedure is described in Algorithm 1. First, the user's location coordinate information is used to match the map data and obtain the user's current road (Step 2). Second, according to $MR_{\mathrm{map}}$, a list of MEC servers that provide services within the current road range is obtained (Step 3). Third, Algorithm 1 migration based on multitier framework selects the appropriate MEC based on the user's movement speed, direction, and task delay attributes (Step 4). Then, the container image and memory snapshot file are synchronized to the selected MEC server (Steps 5–8), and the dirty memory data are iteratively synchronized according to the threshold until the handoff is triggered to stop the container (Steps 9–11). Handoff is a signal for base station switching, and in this paper, we assume that it can be obtained directly from the system. Finally, the container image and the latest runtime file are used to resume the container on the selected MEC server and continue providing services (Step 12). The time complexity of Algorithm 1 is $O(M)$, which is clearly more efficient than the prediction method based on MDP.

## 5. Evaluation

In this section, the performances of the service migration framework described in Section 3 are evaluated. First, the experimental scenario is introduced. Second, the experimental comparison algorithms are presented. Finally, experiments are performed based on the simulation platform and the physical platform.

*5.1. Simulation Scenario.* Our experiment is performed using real-world Beijing map data downloaded from OpenStreetMap [33] to evaluate MSMF. The following are assumed. The local MEC servers and the base stations are placed together, and the regional MEC servers are connected to multiple base stations. The base station placement is hexagonal like in [24] with a 500 m cell separation and NBS = 1951 cells (base stations) in the central area of Beijing City, as shown in Figure 6.

*5.2. Experiment Algorithms.* Representative situations are considered benchmarks to evaluate our method. No matter where the mobile user moves, the service is always migrated to the nearest MEC server for execution of the "Nearest" strategy. Furthermore, our four algorithms are described as follows:

  (i) PM: this algorithm is based on the proposed map-based prediction method for service migration

 (ii) PM-OP: this algorithm is based on the proposed map-based prediction method for service migration, but it is optimized according to $\mathrm{Dis}_j$ when selecting the MEC server

(iii) PM-Tier: this algorithm is based on the proposed map-based prediction method and the multitier

```
Input: MR_map, R_map, P_i, D_i, V_i, L_i, M_d, M_handoff, d_regional, d_local, d_p, s_handoff
Output: Service downtime S and network traffic size N
1: Initialize the algorithm
2: c = matchRoad(R_map, P_i)
3: list = getListMECServers(MR_map, c)
4: m = getNextMECServer(list, D_i, V_i, d_regional, d_local, L_i)
5: Synchronize the image file to m
6: N + = getFileSize(image file)
7: Synchronize the checkpoint files to m
8: N + = getFileSize(checkpoint files)
9: Iteratively synchronize dirty memory data to m based on the threshold M_d until the handoff signal is received
10: N + = getFileSize(total dirty memory data)
11: Stop container and synchronize latest dirty memory data M_handoff to m
12: Restore container on m
13: S = getTime(restore container) − getTime(Stop container)
```

ALGORITHM 1: Migration based on multitier framework.

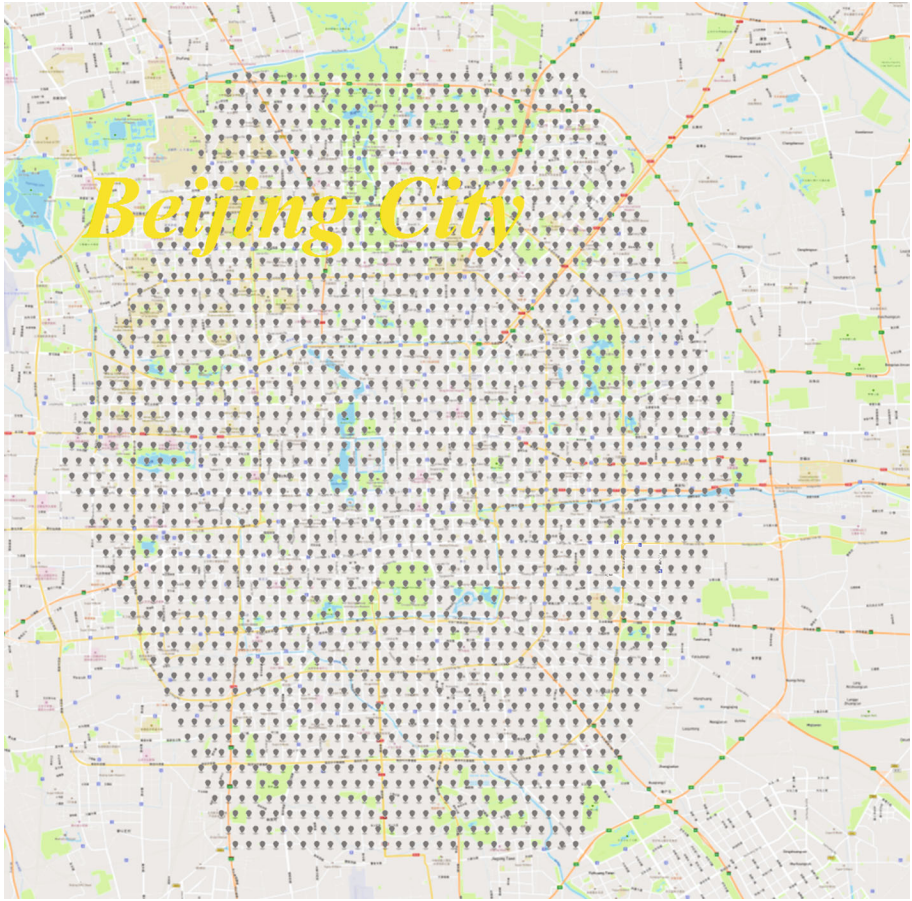

FIGURE 6: Hexagonal base station placement in Beijing Central.

MEC server deployment architecture for service migration

(iv) PM-OP-Tier: this algorithm is based on the proposed map-based prediction method and the multitier MEC server deployment architecture for service migration, but it is optimized according to $Dis_j$ when selecting the migrated MEC server

### 5.3. Simulation-Based Evaluation

*5.3.1. Experiment Setup.* SUMO [34] is used to conduct a system simulation, where users move along the roads or streets in Beijing, China. SUMO (Simulated Urban Traffic) is an open-source multimode traffic simulation software package suitable for handling large networks. SUMO can build traffic system models including road vehicles, public transportation, pedestrians, and real-time access to relevant status data (i.e., position and speed).

In the experiment, Beijing map data are downloaded from OpenStreetMap and loaded into SUMO for simulation experiments. SUMO's tool "radomTrips.py" is used to generate random vehicle trajectories based on Beijing map data. Different vehicle speeds are simulated to evaluate our framework, and the major parameters of the simulation are presented in Table 2. Each experiment sets a different speed for the vehicle, assigns a container of different sizes to each vehicle, runs according to the generated motion trajectory, and then calculates the average of the experimental results. The time to start the container is affected by the size of the container. The larger the container is, the longer the startup time.

All experiment service migrations are based on a precopy method. Each user's task has an attribute level, which identifies the maximum latency that the task can tolerate and determines whether the task will be migrated to a local MEC or regional MEC based on this attribute system.

*5.3.2. Results.* The user's moving speed and the deployment mode of MEC servers greatly influence service migration. The experiment mainly tests the network traffic generated during the service migration and the service downtime time during the service handoff.

*Average network traffic size.* Figure 7(a) shows that the amount of network traffic generated during service migration is related to the speed of user movement and the location of the MEC server (local or regional). The amount of network traffic generated by the "Nearest," "PM," and "PM-OP" methods decreases as the speed increases. When the user moves fast, the user will have left the current MEC service range before the service migration is completed, and only part of the data is copied during this service migration due to the limited coverage of each base station. The deployment method based on a multilayer architecture overcomes such problems caused by the excessively fast movement of users. Moreover, the network traffic generated during service migration is greatly reduced, and the network traffic is reduced by more than 70% to satisfy the delay of user offloading tasks. The two algorithms, "PM-OP" and "PM-OP-Tier," are improved based on the optimization algorithm, which reduces the number of service migrations. Thus, the network traffic generated is reduced accordingly.

*Average service downtime.* Figure 7(b) shows the comparison of service downtime generated by different algorithms at different user movement speeds. When the user is moving at a very low speed, time is sufficient to synchronize the stateless data of the source and target MEC servers, such that the service downtime produced by each algorithm is very close. However, when the moving speed increases, the downtime produced by different algorithms changes considerably, especially the "Nearest" algorithm, because the next access point can be determined only when the user is about to enter the next MEC service range. The faster the speed is, the shorter the time to synchronize data. When switching occurs, a large amount of stateless data has not been synchronized, and a long downtime will be generated. In the optimized method, MEC servers with small coverage areas are removed. When users move into the coverage area of such base stations, service migration is not performed, but the tasks are transferred back to the previous MEC server. This situation can avoid the situation where the service coverage is small, but the moving speed is very fast, preventing the stateless data from being synchronized before the handoff and resulting in a large downtime. The "PM-OP-Tier" algorithm is based on multitier architecture and optimized to select the best MEC server for service migration in real time according to the user's status and the task's attributes. Thus, it is not affected by the user's movement speed.

Table 2: Simulation parameters.

| Parameter | Value |
| --- | --- |
| Coverage for Beijing City | 422 km$^2$ |
| Total number of vehicles | 100 |
| Bandwidth for downlink/uplink | 100 Mb |
| Network latency between UE and local MEC | 10 ms |
| Network latency between UE and regional MEC | 50 ms |
| Number of base station | 1951 |
| Base station coverage radius | 250 m |
| Container size | [100, 2000] MB |
| Container RAM size | [10, 200] MB |
| Dirty memory change rate | 1%/s |
| Threshold for dirty memory synchronization | 40 Mb |
| Container start time | [200, 1000] ms |
| Speed of UEs | [5, 100] km/h |

*5.4. Evaluation on a Real Testbed.* An experiment is conducted using a real testbed and real-world driver traces in this subsection.

*5.4.1. Experiment Setup.* The overall testbed comprises four edge servers, one cloud server, and one end device. Experiments are run in a set of servers with six VMs, in six "host" VMs. Two host VMs act as local MEC servers, two host VMs act as regional MEC servers, one host VM acts as a cloud server, and one host VM acts as a UE. Each host VM has two virtual CPU cores and 2 GB of virtual memory on a physical machine with 2.2 GHz AMD Opteron (TM) Processor 6274 and 64 GB 1600 MHz DDR3 memory. Each edge VM's operating system is Ubuntu 18.04 LTS. Both run (i) runC 1.0.1 as container runtime, (ii) Rsync 3.2.3 as file transfer mechanism, and (iii) CRIU 3.12 for the checkpoint and restore functionality. Linux traffic control [35] is used to control the delay between users, servers, and the cloud center. The bandwidth is configured as 100 Mb/s. The connection delay between the local MEC server and the user is configured as 20 ms, and the connection delay between the regional MEC server and the user is configured as 100 ms. The following applications are used to study our proposed migration framework MSMF.

*Game server.* Xonotic (approximately 1.4 G for LXC) is an addictive arena-style first person shooter game with clear movement and a wide range of weapons. The server part of the game is run in a container, and multiple clients can connect to this part to play together. In our experiment, the client runs on the desktop, while the server and its container live migrate between the MEC servers. This application has a moderate memory requirement of approximately 160 MB.

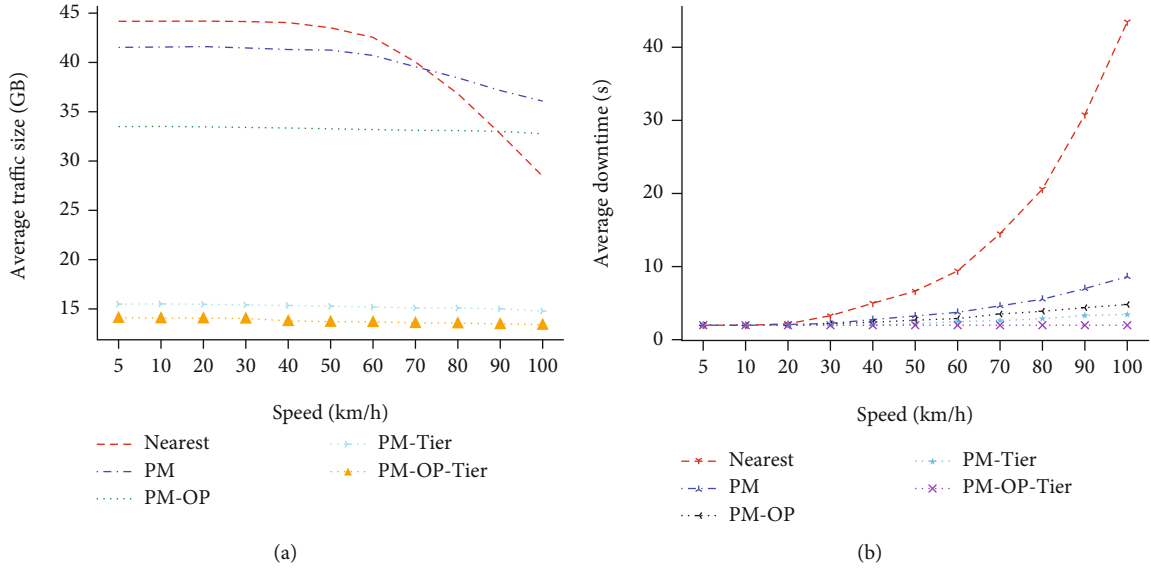

Figure 7: (a) Average network traffic generated by service migration with the different migration strategies. (b) Average service downtime during service migration.

*Face detection.* This application uses the OpenCV library to detect human faces in the video and then sends the results back to the user. The container size of this application is approximately 700 MB, and the size of the required memory is approximately 100 MB.

*RAM simulation.* A script is used to simulate memory-intensive applications, such as Graph computing and deep neural network learning. A container has a simple Python script that consumes a large amount of RAM, and the RAM contents keep changing over time. In our experiments, RAM utilization remains at approximately 300 MB.

First, the real-world GPS trajectory dataset GeoLife [36] is collected in the (Microsoft Research Asia) GeoLife project by 182 users in a period of over three years (from April 2007 to August 2012). This dataset records a broad range of users' outdoor movements with different transportation modes, such as driving, taking a bus, riding a bike, and walking. Ten movement trajectories with different transportation modes (i.e., walk, bikes, bus, and car) are selected for the 10 UEs. A script is run on the VM acting as the UE, sending each user's GPS data separately to the cloud center VM. Second, Barefoot [37] is run on the cloud center VM, and road matching is performed based on the Beijing map data downloaded from OpenStreetMap. Barefoot consists of a software library and a (Docker-based) map server that provides access to street map data from OpenStreetMap and is flexible for use in distributed cloud infrastructure as a map data server or side by side with Barefoot's standalone server for offline (matcher server) and online map matching (tracker server).

Third, a test application is selected for each experiment, an appropriate edge server for service migration testing is selected according to the application's latency level, and the average of the experimental results is calculated.

*5.4.2. Results.* The network traffic generated during service migration is directly related to the type of application, and the downtime during service handoff is related to the algorithm used by users and the moving speed.

*Average network traffic size.* The size of network traffic generated by service migration in a MEC environment is directly related to the size of the container image used and the type of application, as shown in Figure 8(a). Data migrated in the service migration mainly includes container image and memory data. The sizes of memory data generated by different applications are different. Therefore, the network traffic generated by service migration is affected by the size of memory data generated by applications in addition to the size of the image, as shown in Figure 8(b).

Applications with a large memory footprint generate large amounts of data during service migration.

*Average service downtime.* Figure 8(c) shows the service downtime generated by service migration under different traffic modes with different algorithms. It does not make a difference when the moving speed is very low, but it will make a large difference when selecting different means of transportation, especially "Nearest." When the user is walking, the downtime generated by the "Nearest" algorithm is close to that of other algorithms. The algorithm based on multilayer architecture is clearly superior to other algorithms, especially for the traffic mode with fast-moving speed. The optimized algorithm also has good performance. The best performance is the optimized algorithm based on a multitier architecture. It can handoff seamlessly between different modes without being affected by the mobile mode, providing users with more stable services.

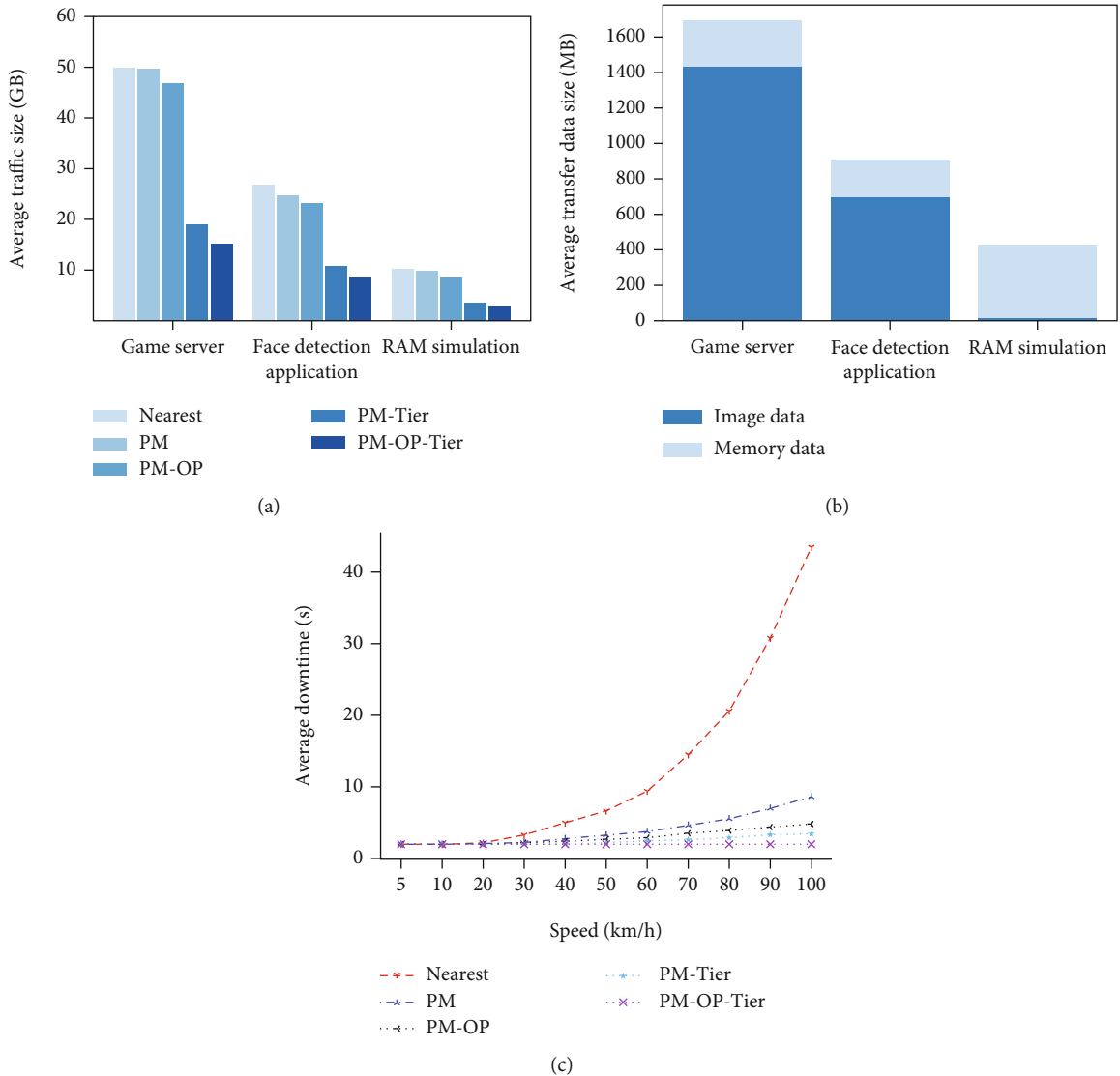

Figure 8: (a) Average network traffic. (b) Average data size for each migration with PM-OP-Tier. (c) Average service downtime.

## 6. Conclusion

In this paper, the service migration problem in MEC is studied based on map prediction. A novel multilayer MEC server deployment framework is designed based on map data online prediction. The interruption time and network redundant traffic generated during service migration can be reduced as much as possible to meet the delay requirements of offloading tasks. Considering that the trajectory of individuals in real life is not random and given several restrictions (i.e., buildings and parks), individuals will walk along a given road. The mapping relationship between the deployment location of the MEC server and the road based on digital map data is built, and real-time prediction is performed by collecting the user's movement status (i.e., position, direction, and acceleration) in real time to deal with the problem of real-time prediction of the next MEC server that the user will access. A framework is designed based on multitier deployment of MEC servers considering the differences in user movement speed and offloading task delay requirements to reduce network traffic and service interruptions caused by service migration and to improve QoS and QoE. The results demonstrate the superior performance of our method through extensive experiments. Future research directions include the privacy protection of user data and the analysis of the use of the runtime environment. The user's private data include task offloading data and location data. MEC servers retain frequently used containers in different time periods instead of deleting them after use, which can reduce the redundant traffic generated by service migration and improve service quality.

## Data Availability

We create a simulation to demonstrate the performance of our proposed algorithm.

## Conflicts of Interest

## Acknowledgments

The work was supported by the National Key R&D Program of China under grant no. 2020YFB1406902 and the National Natural Science Foundation of China (NSFC) under grant nos. 61672186 and 61872110.

## References

[1] L. Yang, L. Bo, J. Cao, Y. Sahni, and Z. Wang, "Joint computation partitioning and resource allocation for latency sensitive applications in mobile edge clouds," in *2017 IEEE 10th International Conference on Cloud Computing (CLOUD)*, Honolulu, HI, USA, 2017.

[2] Y. C. Hu, M. Patel, D. Sabella, N. Sprecher, and V. Young, "Mobile edge computing—a key technology towards 5G," *ETSI White Paper*, vol. 11, no. 11, pp. 1–16, 2015.

[3] K. Peng, V. C. M. Leung, X. Xu, L. Zheng, J. Wang, and Q. Huang, "A survey on mobile edge computing: focusing on service adoption and provision," *Wireless Communications and Mobile Computing*, vol. 2018, Article ID 8267838, 16 pages, 2018.

[4] P. Mach and Z. Becvar, "Mobile edge computing: a survey on architecture and computation offloading," *IEEE Communications Surveys & Tutorials*, vol. 19, no. 3, pp. 1628–1656, 2017.

[5] I. A. Elgendy, W.-Z. Zhang, C.-Y. Liu, and C.-H. Hsu, "An efficient and secured framework for mobile cloud computing," *IEEE Transactions on Cloud Computing*, vol. 9, no. 1, pp. 79–87, 2021.

[6] Y. Mao, C. You, J. Zhang, K. Huang, and K. B. Letaief, "A survey on mobile edge computing: the communication perspective," *IEEE Communications Surveys & Tutorials*, vol. 19, no. 4, pp. 2322–2358, 2017.

[7] C.-H. Hsu, S. Wang, Y. Zhang, and A. Kobusinska, "Mobile edge computing," *Wireless Communications and Mobile Computing*, vol. 2018, 3 pages, 2018.

[8] I. A. Elgendy, W. Z. Zhang, Y. Zeng, H. He, Y.-C. Tian, and Y. Yang, "Efficient and secure multi-user multitask computation offloading for mobile-edge computing in mobile IoT networks," *IEEE Transactions on Network and Service Management*, vol. 17, no. 4, pp. 2410–2422, 2020.

[9] G. Li, J. Wang, J. Wu, and J. Song, "Data processing delay optimization in mobile edge computing," *Wireless Communications and Mobile Computing*, vol. 2018, Article ID 6897523, 9 pages, 2018.

[10] R. Yadav and W. Zhang, "MeReg: managing energy-SLA tradeoff for green mobile cloud computing," *Wireless Communications Mobile Computing*, vol. 2017, no. 2, pp. 1–11, 2017.

[11] M. Satyanarayanan, G. A. Lewis, E. J. Morris, S. Simanta, J. Boleng, and K. Ha, "The role of cloudlets in hostile environments," *IEEE Pervasive Computing*, vol. 12, no. 4, pp. 40–49, 2013.

[12] A. Machen, S. Wang, K. K. Leung, B. Ko, and T. Salonidis, "Live service migration in mobile edge clouds," *IEEE Wireless Communications*, vol. 25, no. 1, pp. 140–147, 2018.

[13] F. Bonomi, R. A. Milito, J. Zhu, and S. Addepalli, "Fog computing and its role in the internet of things," in *Proceedings of the First Edition of the MCC Workshop on Mobile Cloud Computing - MCC '12*, pp. 13–16, Helsinki, Finland, 2012.

[14] T. Taleb and A. Ksentini, "Follow me cloud: interworking federated clouds and distributed mobile networks," *IEEE Network*, vol. 27, no. 5, pp. 12–19, 2013.

[15] S. Wang, J. Xu, N. Zhang, and Y. Liu, "A survey on service migration in mobile edge computing," *IEEE Access*, vol. 6, pp. 23511–23528, 2018.

[16] S. Fu, Z. Su, Y. Jia et al., "Interference cooperation via distributed game in 5g networks," *IEEE Internet of Things Journal*, vol. 6, no. 1, pp. 311–320, 2019.

[17] K. Ha, Y. Abe, Z. Chen et al., *Adaptive VM Handoff across Cloudlets*, Technical Report CMU-CS-15-113, 2015.

[18] L. Ma, S. Yi, N. Carter, and Q. Li, "Efficient live migration of edge services leveraging container layered storage," *IEEE Transactions on Mobile Computing*, vol. 18, no. 9, pp. 2020–2033, 2019.

[19] J. Plachy, Z. Becvar, and E. C. Strinati, "Dynamic resource allocation exploiting mobility prediction in mobile edge computing," in *2016 IEEE 27th Annual International Symposium on Personal, Indoor, and Mobile Radio Communications (PIMRC)*, pp. 1–6, Valencia, Spain, 2016.

[20] A. Ksentini, T. Taleb, and M. Chen, "A Markov decision process-based service migration procedure for follow me cloud," in *2014 IEEE International Conference on Communications (ICC)*, Sydney, NSW, Australia, 2014.

[21] S. Wang, R. Urgaonkar, T. He, M. Zafer, K. Chan, and K. K. Leung, "Mobility-induced service migration in mobile microclouds," in *2014 IEEE Military Communications Conference*, Baltimore, MD, USA, 2015.

[22] S. Wang, R. Urgaonkar, M. Zafer, T. He, K. S. Chan, and K. K. Leung, "Dynamic service migration in mobile edge-clouds," in *2015 IFIP Networking Conference (IFIP Networking)*, pp. 1–9, Toulouse, France, 2015.

[23] S. Wang, R. Urgaonkar, T. He, K. Chan, M. Zafer, and K. K. Leung, "Dynamic service placement for mobile micro-clouds with predicted future costs," *IEEE Transactions on Parallel and Distributed Systems*, vol. 28, no. 4, pp. 1002–1016, 2017.

[24] S. Wang, R. Urgaonkar, M. Zafer, T. He, K. Chan, and K. K. Leung, "Dynamic service migration in mobile edge computing based on Markov decision process," *IEEE/ACM Transactions on Networking*, vol. 27, no. 3, pp. 1272–1288, 2019.

[25] Q. Wu, X. Chen, Z. Zhou, and L. Chen, "Mobile social data learning for user-centric location prediction with application in mobile edge service migration," *IEEE Internet of Things Journal*, vol. 6, no. 5, pp. 7737–7747, 2019.

[26] Y.-Y. Chiang, S. Leyk, and C. A. Knoblock, "A survey of digital map processing techniques," *ACM Computing Surveys*, vol. 47, no. 1, pp. 1–44, 2014.

[27] V. Sharma, D. N. K. Jayakody, and M. Qaraqe, "Osmotic computing-based service migration and resource scheduling in mobile augmented reality networks (MARN)," *Future Generation Computer Systems*, vol. 102, pp. 723–737, 2020.

[28] F. Zhang, G. Liu, F. Xiaoming, and R. Yahyapour, "A survey on virtual machine migration: challenges, techniques, and open issues," *IEEE Communications Surveys & Tutorials*, vol. 20, no. 2, pp. 1206–1243, 2018.

[29] "CRIU," https://criu.org/Main_Page.

[30] P. HaulAugust 2020, https://criu.org/P.Haul.

[31] J. Hu, G. Wang, X. Xu, and Y. Lu, "Study on dynamic service migration strategy with energy optimization in mobile edge computing," *Mobile Information Systems*, vol. 2019, Article ID 5794870, 12 pages, 2019.

[32] T. Taleb and A. Ksentini, "An analytical model for follow me cloud," in *2013 IEEE Global Communications Conference (GLOBECOM)*, pp. 1291–1296, Atlanta, GA, USA, 2013.

[33] OpenStreetMap Wiki, "Main page — openstreetmap wiki," 2020, August 2020 https://wiki.openstreetmap.org/w/index.php?title=Main_Page&oldid=2013332.

[34] P. A. Lopez, E. Wiessner, M. Behrisch et al., "Microscopic traffic simulation using sumo," in *In The 21st IEEE International Conference on Intelligent Transportation Systems*, Maui, HI, USA, 2018.

[35] M. A. Brown, "Traffic control howto. [EB/OL]," April 2020, https://tldp.org/HOWTO/Traffic-Control-HOWTO/.

[36] Y. Zheng, X. Xie, and W.-Y. Ma, "Geolife: a collaborative social networking service among user, location and trajectory," *IEEE Data Engineering Bulletin*, vol. 33, no. 2, pp. 32–39, 2010.

[37] "Barefoot," August 2020, https://github.com/bmwcarit/barefoot/wikiaccessed..